\documentclass[10pt]{article}
\usepackage[letterpaper]{geometry}
\usepackage{hicss}
\usepackage{times}
\usepackage[none]{hyphenat}
\usepackage{url}
\usepackage{latexsym}
\usepackage[frozencache,cachedir=.]{minted}
\usepackage{indentfirst}
\usepackage{graphicx}
\graphicspath{{images/}}
\usepackage[style=apa]{biblatex}
\title{{V}is{P}ile: A Visual Analytics System for Analyzing Multiple Text Documents With Large Language Models and Knowledge Graphs}

\author{Adam Coscia \\
 Georgia Institute of Technology \\
 {\underline{ acoscia6@gatech.edu}} \\ \And
 Alex Endert \\
 Georgia Institute of Technology \\
 {\underline{ endert@gatech.edu} } }

\date{}

\newif\ifauthornotes
\newif\ifstrike
\newif\iftodo
\newif\ifrevise
\newif\ifadd
\newif\ifreplace

\newcommand{\add}[1]{\ifadd{\leavevmode\color{magenta}{#1}}\else{#1}\fi}

\begin{document}

\maketitle

\begin{abstract}
Intelligence analysts perform sensemaking over collections of documents using various visual and analytic techniques to gain insights from large amounts of text.
As data scales grow, our work explores how to leverage two AI technologies, large language models (LLMs) and knowledge graphs (KGs), in a visual text analysis tool, enhancing sensemaking and helping analysts keep pace.
Collaborating with intelligence community experts, we developed a visual analytics system called \textbf{VisPile}.
VisPile integrates an LLM and a KG into various UI functions that assist analysts in grouping documents into piles, performing sensemaking tasks like summarization and relationship mapping on piles, and validating LLM- and KG-generated evidence.
Our paper describes the tool, as well as feedback received from six professional intelligence analysts that used VisPile to analyze a text document corpus.
\end{abstract}

\subsubsection*{Keywords:}

Visual analytics, Sensemaking, Text analytics, Large language models, Knowledge graphs.

\section{Introduction}
\label{sec:introduction}

Performing visual text analysis is an ongoing challenge in the visual analytics community, where reading, understanding, and making sense of large amounts of text is time-consuming.
As data scales increase, decision-makers in the intelligence community seek new methods to balance the constant tension between performing human and automated analysis (\cite{United:2019:AIM, Sayler:2020:AINatSec}).
Automated methods, including AI, can help extract and summarize relevant text and reduce the need for extensive reading, while human-in-the-loop interfaces help people combine visual and analytical methods to contextualize evidence and synthesize critical insights (\cite{Baker:2009:VisEnhancesSensemaking}).

In visual analytics, two recent technologies have dominated the discourse: large language models (LLMs) and knowledge graphs (KGs).
LLMs can rapidly summarize text, explain concepts, and answer questions (\cite{Srivastava:2023:BeyondImitationGame}).
KGs can validate ground-truth information generated by LLMs and link entities and events back to source text (\cite{Li:2024:KGChallenges}).
Recent visual analytics tools are exploring how LLMs and KGs can assist in task planning, question-answering, and summarization of unstructured text.
LLMs and KGs could help analysts distill large document sets into relevant data and synthesize new evidence, while minimizing the need for detailed manual reading.
However, understanding how to effectively integrate these technologies into visual text analytic tools is less well understood.

In this work, we explore how LLMs and KGs can be integrated into visual text analysis through a yearlong design study with domain experts in the intelligence community.
While LLMs and KGs show promise in automating time-consuming analysis tasks, it is unclear how to design a visual analytics tool around their emergent capabilities.
To address these gaps, we present \textbf{VisPile} (Sect.~\ref{sec:system}), a visual analytics system for analyzing text documents.
Our system incorporates LLM and KG features to help analysts group documents into piles and perform sensemaking tasks (summarization, extraction, Q\&A, etc.) on piles.
We demonstrate how LLM and KG features in VisPile impact sensemaking through formative feedback from six professional intelligence analysts (Sect.~\ref{sec:expert_feedback}).
The analysts used the LLM and KG to quickly find and compare relevant subsets of $845$ documents for piling, chained LLM tasks and KG facts together to enable deeper analysis of their piles, and combined the LLM response and KG suggestions to contextualize evidence and uncover hidden connections.

In summary, we contribute: (1) design goals for integrating LLMs and KGs into visual text analysis; (2) \textbf{VisPile}, an open-source\footnote{VisPile code: \url{https://github.com/AdamCoscia/VisPile}} visual analytics tool with LLM and KG features for text document analysis; (3) domain expert feedback illustrating preliminary results on how LLMs and KGs impact sensemaking in text analysis.

\section{Background and Related Work}
\label{sec:related_work}

We first describe the intelligence analysis process and related visual analytics systems, then recent usage of LLMs and KGs in visual text analysis.

\subsection{Visual Analytics for Intelligence Analysis}
\label{sec:related_intel_process}
Our designs are grounded in the Pirolli and Card sensemaking loop for intelligence analysis (\cite{Pirolli:2005:SensemakingProcess}).
In a bottom-up sensemaking process, intelligence analysts iteratively schematize information spread across a corpus of text documents like news articles.
To gain situational awareness, analysts usually group related documents into piles, grounding their understanding of both relevant and irrelevant information (\cite{Shipman:1999:IncrementalFormalism}).
With piles of documents at hand, analysts read through them to gather reportable evidence, usually snippets describing key people or events related to the pile’s topic.
Finally, raw evidence needs to be verified, typically by mapping it to documents in and across piles.

Research in visual analytics has a rich history of developing techniques for sensemaking over unstructured text data like documents (\cite{Mccolgin:2006:QAtoVA, Kang:2011:VAIntelProcess, Endert:2014:HumanIsTheLoop}).
For example, interactive topic modeling (\cite{Berry:2010:TextMining}) lets users guide ML models to identify documents relevant to bottom-up sensemaking.
Tools like TopicSifter (\cite{Kim:2019:TopicSifter}) enable human-in-the-loop feedback by allowing users to target documents to guide the search process.
Spatial analysis methods have also proven effective for text document sensemaking (\cite{Andrews:2010:SpaceToThink, Endert:2012:SemanticInteraction}).
Tools like Analyst's Notebook (\cite{i2:2024:AnalystNotebook}) support evidence marshalling by organizing documents into related groups.
Other systems span multiple stages of sensemaking; e.g., Jigsaw (\cite{Stasko:2007:Jigsaw}) uses multiple coordinated views to reveal relationships across documents.
\add{Recent works have begun leveraging LLMs and KGs in visual analytics. KNowNEt (\cite{Yan:2025:KNowNEt}) uses a KG to enhance LLM-driven question-answering but does not support multi-document analysis or evidence marshalling. LEVA (\cite{Zhao:2025:LEVA}) augments visual analytics systems with automatic insights and report generation but lacks controls for selecting sensemaking tasks.}

\subsection{LLMs \& KGs For Visual Text Analysis}
\label{sec:related_llms_kg}
LLMs and KGs have individually demonstrated impressive data analysis capabilities, inspiring us to consider their potential to support the intelligence analysis process.
Specifically, we focus on the potential for LLMs and KGs to help analysts retrieve relevant text documents and perform sensemaking over them.

\add{LLMs demonstrate comparable performance to humans in fetching, parsing, and visualizing tabular data autonomously (\cite{Cheng:2023:GPT4DataAnalyst, Zhang:2024:DataCopilot}).}
\add{However, the space of text document analysis is less explored.}
Classification models like BERTopic (\cite{Grootendorst:2022:BERTopic}) can generate open-ended topic models from a corpus of text documents, providing an entry point into exploring the document space when little is known up-front.
Alternatively, generative models like OpenAI's GPT-4o can perform a semantic similarity search using retrieval-augmented generation (RAG)
techniques (\cite{Lewis:2020:RAGLLM}) to return relevant documents based on an open-ended query.
Given a group of documents, LLMs can perform tasks like entity extraction, summarization, and question-answering using prompts to guide the LLM's output (\cite{Liu:2023:Prompting}).
LLMs could help analysts shift time spent away from reading and towards synthesizing information.

Conversely, KGs are a more nascent technology in the text analysis space.
KGs encode semantic relationships between entities as triples in the form subject$\to$predicate$\to$object, with metadata on the nodes (subject, object) and edges (predicate) as properties (\cite{Li:2024:KGChallenges}).
Consider the triple John$\to$likes$\to$Sally, where John and Sally might be labeled as Person type and likes as Emotion type in their metadata.
Their semantic nature makes them particularly suited for providing context to text content, e.g., text generated by an LLM (\cite{Hogan:2021:KGs}).
Most prior KG systems focus on direct question-answering (\cite{Li:2024:LinkQ}) or visual exploration of existing graph structures like Wikipedia (\cite{Latif:2021:VisKonnect}).
However, recent work has explored using LLMs to extract triples and create a KG directly from a document corpus (\cite{Pan:2024:UnifyingLLMsKGs}).
We utilize this technique in our work to open the door for exploring new text analysis methods using the KG.

\section{The {VisPile} System}
\label{sec:system}

We present \textbf{VisPile}, a visual analytics tool for text document analysis.
VisPile integrates an LLM and a KG into searching, filtering, and piling documents, analyzing documents in piles, and validating LLM- and KG-generated evidence.
Users combine LLM and KG features in VisPile into various sensemaking workflows.

Throughout our system description and expert evaluation, we used the IEEE 2014 VAST Challenge dataset (KRONOS) as a proof-of-concept (\cite{Whiting:2014:VAST2014Challenge}).
The dataset includes 845 plain-text news reports (500–1000 words each) describing complex relationships that culminated in a kidnapping on the fictitious island nation of Kronos.
The central task is: \textit{``Synthesize potential relationships that may have led to the disappearance.''}
The KRONOS dataset is widely used as an unclassified proxy for intelligence analysis.
We also used OpenAI’s GPT-3.5 Turbo, the state-of-the-art generative LLM at the time, to process data and run prompts.
Our data architecture supports any similarly structured dataset and generative LLM, including open-source models.
All prompts are available in the repository and supplemental material.

\subsection{Design Goals}
\label{sec:design_goals}
To develop VisPile, we followed the methodology for visualization design studies proposed by \textcite{Sedlmair:2012:DesignStudyMethods}.
Our team comprised\add{ the authors,} visualization experts in human-centered computing and data analysis,\add{ working with} professional SIGINT analysts, program coordinators, and managers with years of combined experience working in the U.S. Intelligence Community (IC).
We first identified where LLMs and KGs could enhance stages of sensemaking in intelligence analysis.
We then refined what LLMs and KGs would do and how analysts would interact with them through bi-weekly focus groups.
In each session, \add{the visualization experts} presented low-fidelity mockups to gather and refine iterative design feedback.
Co-designing with domain experts \add{in both visualization and the IC} helped us synthesize shared design goals towards enhancing the sensemaking process.

\begin{figure*}[!t]
  \centering
  \includegraphics[width=\linewidth]{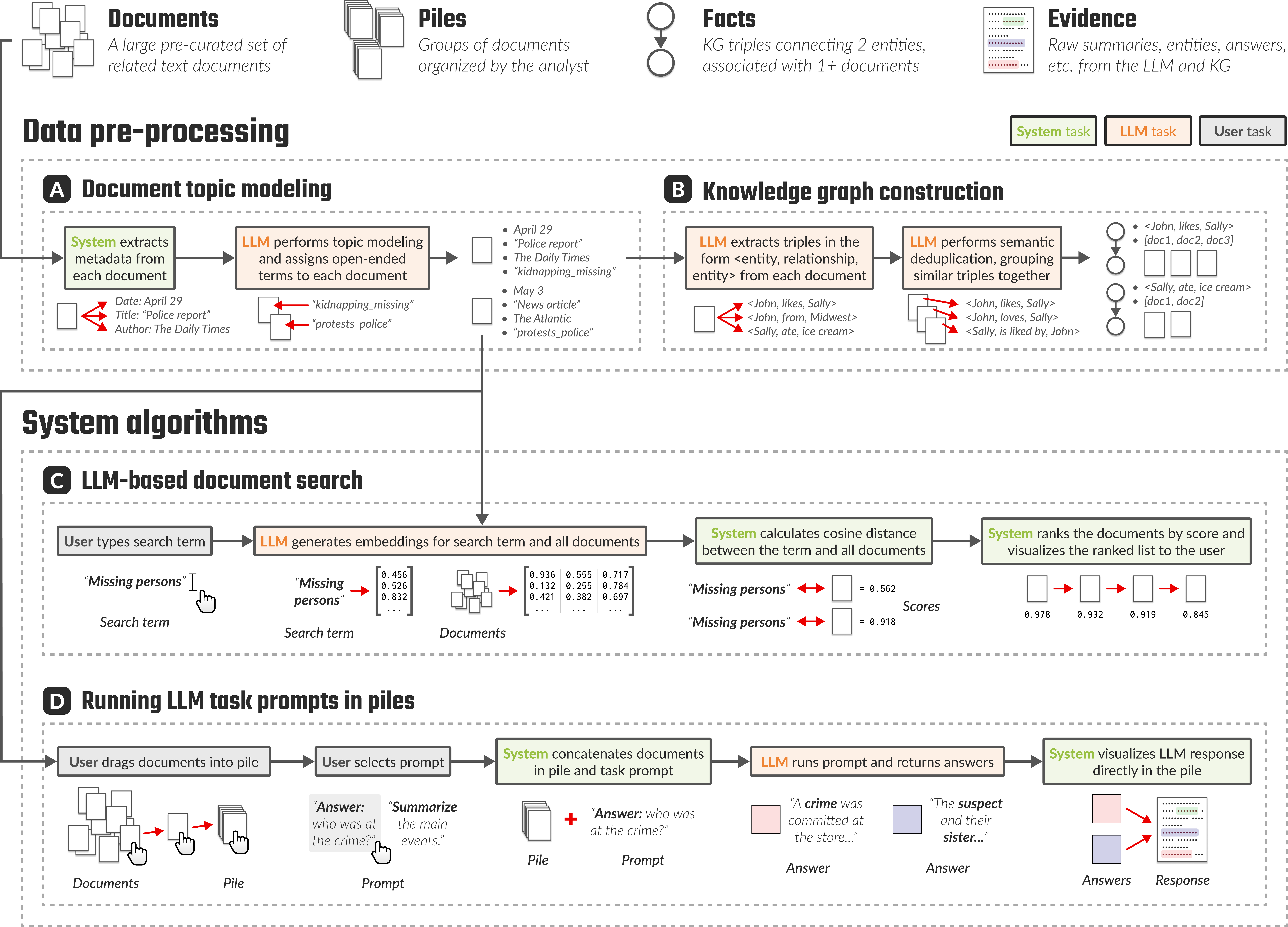}
  \caption{%
  	The data architecture for VisPile.
  }%
  \label{fig:architecture}
\end{figure*}

\add{We established a shared technical terminology (Fig.~\ref{fig:architecture}, top) based on the analysts' workflows, used throughout the paper. To scope our investigation, we focus on analyzing text-only \textbf{documents} like technical reports or news articles. Analysts often organize their reading by manually arranging documents into \textbf{piles}, groups of documents that share a common topic or purpose. As analysts read documents, they often perform entity-based analysis, searching for \textbf{facts} that represent the relationship between two entities; e.g., ``John likes Sally.'' They may also generate intermediate artifacts like summaries or answers to questions as they work, which we refer to collectively as \textbf{evidence}.}

In VisPile, LLMs and KGs are the primary method of finding and grouping relevant documents into meaningful piles, using piles to extract\add{ facts and} evidence from documents, and analyzing\add{ facts and} evidence for insights.
This represents a fundamental shift in the intelligence analysis process, where fact/evidence gathering is traditionally a cognitive process of the analyst only.
The LLM fulfills the role of synthesizing massive amounts of text quickly to generate/answer questions and summarize relevant entities, relationships, topics, facts, etc.
KGs, on the other hand, already represent the distillation of text documents into a set of interconnected facts.
Analysts can traverse facts in the KG directly, only reading documents when necessary, as well as compare LLM-generated evidence with known KG facts. 

Piles are vital for scoping the data that automated functions are allowed to operate on.
VisPile enables direct manipulation of piles in a spatial view, where analysts drag documents directly into different areas of the workspace and arrange the ordering of piles.
The underlying data architecture then handles running LLM and KG functions directly on the documents in piles.
Choosing how best to pile documents and analyze piles for the task at hand promotes analyst agency during sensemaking.
Our design goals (\textbf{G}) each address aspects of supporting different user tasks within this workflow.

\medskip
\noindent\textbf{G1 - Leverage LLM for open-ended document search. }
The LLM should help users forage for documents by recommending open-ended topic groups and interpreting open-ended task descriptions like \textit{``Who are the relevant actors?''} for enabling semantic search.

\medskip
\noindent\textbf{G2 - Link KG data to source documents. }
Node and edge metadata should be displayed that links the graph data to source documents, allowing users to easily identify groups of related source documents and create piles from related KG facts.

\medskip
\noindent\textbf{G3 - Run LLM sensemaking tasks within piles. }
Users should be able to prompt LLMs to perform analytic tasks on multiple documents in piles.
Prompts should be customizable, allowing users to modify and combine operations and even generating custom tasks to target more unique and complex operations.

\medskip
\noindent\textbf{G4 - Traverse connected KG facts related to piles. }
Users should be able to interact with and explore the KG visually, connecting facts to their pile for refining their evidence and gathering new evidence from source documents related to their piles.

\medskip
\noindent\textbf{G5 - Enable validation operations on LLM responses. }
Users should be able to perform validation operations that keep them in the loop of linking statements and sources to verify the veracity of LLM-generated text.
Operations include extracting ground-truth KG data from LLM responses, linking LLM responses to related source documents, and suggesting additional documents relevant to LLM-generated evidence.

\medskip
\noindent\textbf{G6 - Provide related context around KG facts. }
Connected and related nodes should automatically be surfaced when traversing the KG, enabling users to contextualize KG facts based on pile evidence.

\subsection{Data Architecture}
\label{sec:data_architecture}
VisPile operates within a closed document model, in which a fixed set of related documents is pre-curated but too large to read everything (e.g., $\approx1000$ documents).
VisPile takes as input a corpus of plain text documents.
Documents are first pre-processed to log metadata such as title, length, and topic.
Then, a KG is constructed from the document text, consisting of a list of fact triples with metadata including the source(s) where the fact came from.
The list of documents and KG facts are then visualized in a web-based application built with Vue.js.

To generate document topic models (Fig.~\ref{fig:architecture}A), VisPile uses BERTopic (\cite{Grootendorst:2022:BERTopic}), an LLM-based topic modeling approach.
\add{Because LLMs are good at generating open-ended classes, we experimented with not pre-defining topics beforehand and observing how that could improve the piling process.}
BERTopic produces open-ended semantic topics such as \textit{``kidnapping\_missing''} and \textit{``protests\_police''}, which are assigned to each document as metadata.
LLMs are also used to perform sensemaking tasks over groups of documents in piles (Fig.~\ref{fig:architecture}D).
Following \cite{Shaib:2023:LLMsForMultidocAnalysis}, each task prompt uses text separator tokens to combine documents and the task description into a single prompt.
The output from the LLM is then visualized directly in the web app.

\begin{figure*}[!t]
  \centering
  \includegraphics[width=\linewidth]{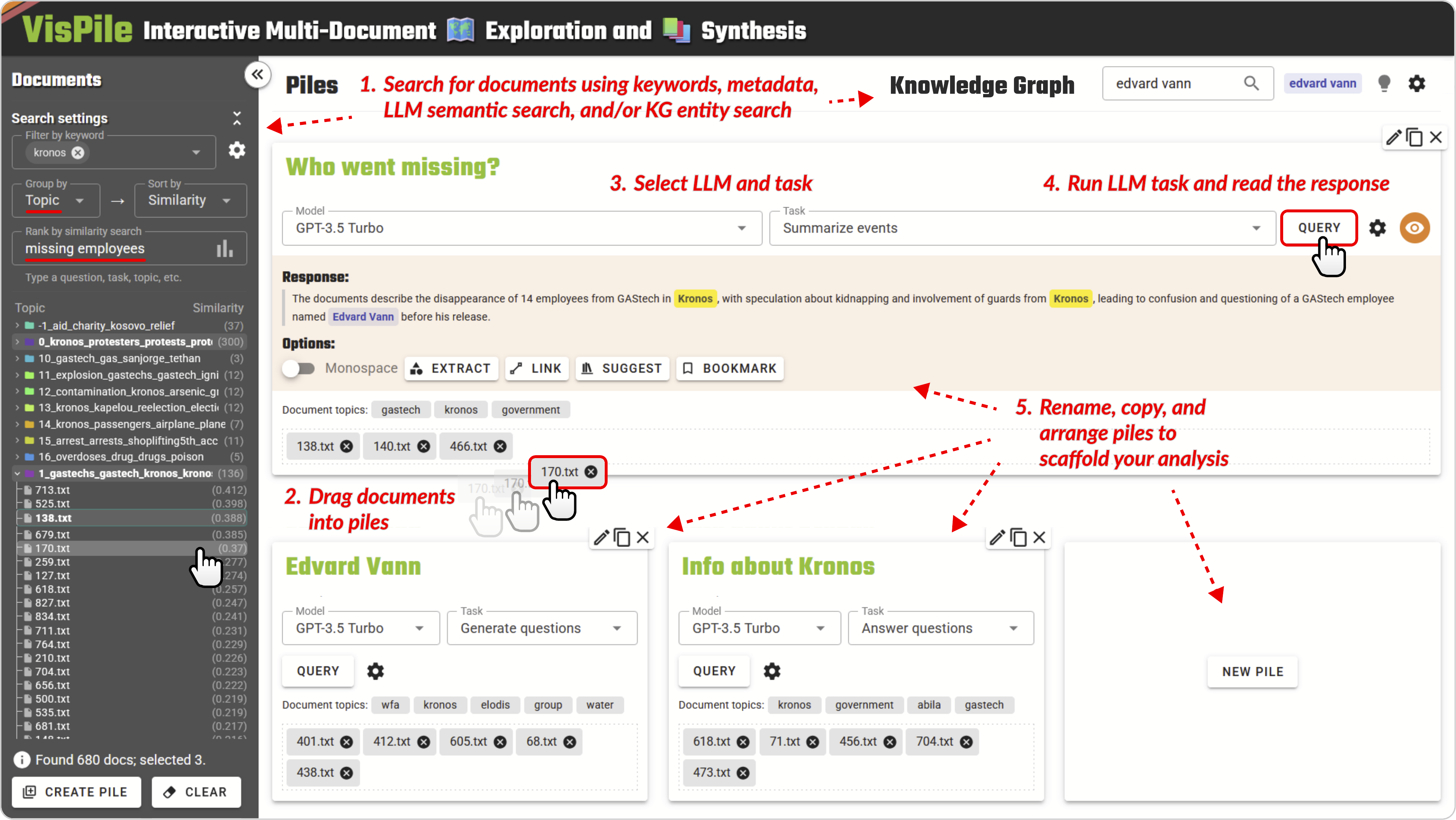}
  \caption{%
    The VisPile interface.
  	VisPile leverages an LLM and a KG to help analysts gather evidence from document collections.
    The general user workflow is: (1) Search for documents to pile, using keywords, metadata, KG entity search, and/or LLM semantic search; (2) drag documents into piles; (3) choose an LLM and pre-generated task to run as an LLM prompt; (4) Run the LLM task and read the response; (5) Repeat the process, rearranging and renaming piles to scaffold the sensemaking process.
  }%
  \label{fig:interface}
\end{figure*}

To generate the KG, we follow the method proposed by \textcite{Pan:2024:UnifyingLLMsKGs} (Fig.~\ref{fig:architecture}B).
\add{Similarly to topic modeling, we wanted to explore the benefits that open-ended, LLM-based KG extraction could provide over traditional, schema-based extraction.}
First, all documents are fed sequentially to a generative LLM (GPT-3.5 Turbo) and raw triples are extracted using prompt engineering (\cite{Sahoo:2024:PromptEngineering}) in the form $<$entity, relationship, entity$>$\add{, representing the style of facts commonly used in entity-based analysis in the IC. We then map facts to the subject$\to$predicate$\to$object form of a KG.}
We save the source document of each triple as metadata.
Then, to validate the extracted facts, we run semantic deduplication by grouping repeated facts based on cosine similarity of entity and relationship embeddings.
From each group, we select a representative fact and record all source documents for the duplicates.
This results in a list of non-overlapping triples, each linked to its supporting documents that can be cited.
The number of supporting documents serves as a proxy for ``support''; i.e., facts cited in more documents may be more trustworthy.

\subsection{System Features}
\label{sec:system_features}
VisPile (Fig.~\ref{fig:interface}) presents several features to support sensemaking: searching and filtering documents using LLM document search and the KG fact list; putting documents into piles and analyzing piles using LLMs and KGs; and validating LLM and KG evidence in piles.

\medskip
\noindent\textbf{LLM-based document search. }
Documents in VisPile are shown in a list (Fig.~\ref{fig:interface}, left) that can be filtered using traditional keyword search, as well as grouped and sorted by document metadata such as date, name, and length.
VisPile enables open-ended document search (\textbf{G1}) in two ways.
First, users can organize documents by LLM-generated topic using BERTopic (Fig.~\ref{fig:architecture}A), leveraging the strengths of the LLM to take a first-pass over organizing documents by relevant topics.
Second, users can perform a retrieval-augmented generation (RAG)-based semantic similarity search (Fig.~\ref{fig:architecture}C).
The user's search query and each document are embedded using a text embedding model (e.g., OpenAI's text-embedding-3).
Document embeddings are then ranked by cosine similarity to the search query.
The ranked list is shown back to the user, enabling a more dynamic and iterative search process.

\begin{figure*}[!t]
  \centering
  \includegraphics[width=\linewidth]{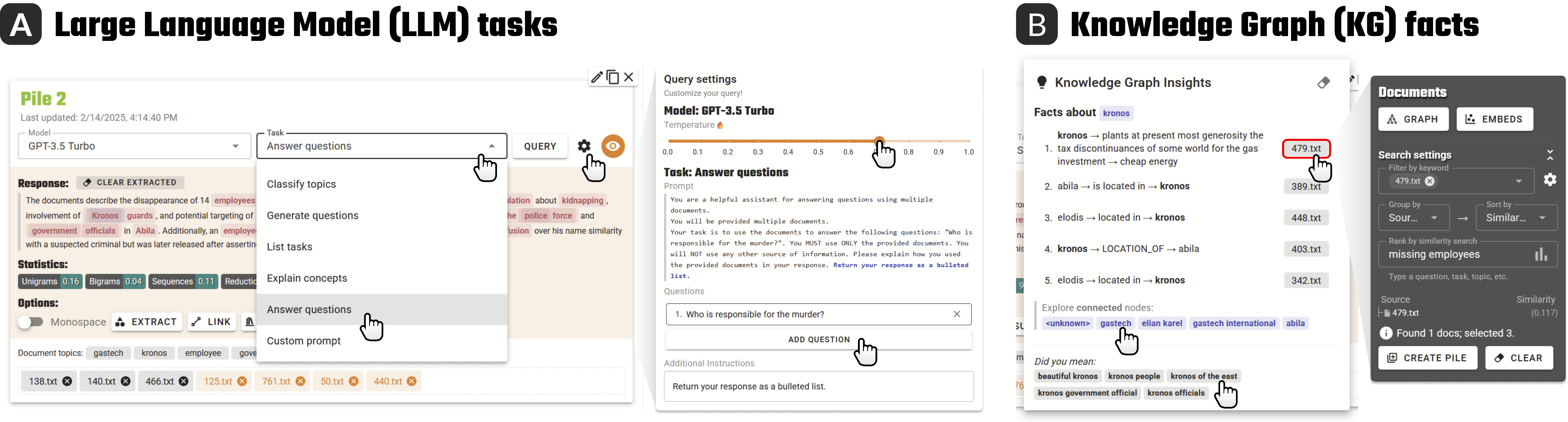}
  \caption{%
    Piles (A) allow users to group documents together and run any of nine sensemaking tasks on the documents as LLM prompts.
    Prompts are shown for transparency and can be adjusted with inputs for questions, entity types, and concepts, as well as a temperature slider.
  	The KG fact list (B) shows up to 5 top-ranked facts, or plain text triples in the form subject$\to$object$\to$predicate, based on the LLM response in the pile.
  }%
  \label{fig:piles_and_facts}
\end{figure*}

\medskip
\noindent\textbf{KG fact list. }
The KG can similarly be used for open-ended document search.
Users can access entities in the KG using a free-text search bar (Fig.~\ref{fig:interface}, top).
Visualizing large, connected KGs using traditional node-link diagrams can quickly become cognitively demanding (\cite{Li:2024:KGChallenges}).
Instead, VisPile represents the resulting graph around a searched entity as a list of facts (Fig.~\ref{fig:piles_and_facts}B).
The KG fact list shows up to 5 facts, or plain text triples in the form subject$\to$object$\to$predicate.
The 5 facts are ranked in the same way as the LLM semantic search (Fig.~\ref{fig:architecture}C), by embedding the pile text and all facts, then pairwise comparing them to get a ranked list of top-scoring facts.
Source documents for each fact are shown to the right of facts, allowing users to search for documents directly using KG facts (\textbf{G2}).
Users can click on a document name to filter for that document in the documents view.

\medskip
\noindent\textbf{Document piles, LLM tasks, and KG traversal. }
Once a user finds documents they want to analyze, they can drag them into a pile (Fig.~\ref{fig:piles_and_facts}A) and rename the pile to track their line of questioning.
They can create as many piles as they want, and each can be renamed, duplicated, sorted and filtered.

Piles have a model selection drop-down and $9$ different sensemaking tasks implemented as LLM prompts (\textbf{G3}).
VisPile breaks down some prompts into adjustable parts to enable customization, including inputs for questions, entity types, and concepts, as well as a slider to adjust the LLM temperature, which controls the LLM's creativity.
Prompts are always shown verbatim to increase transparency.
The prompts were developed through iterative prompt engineering with feedback from our IC collaborators and are available in the supplemental material.
They include:

\begin{enumerate}
    \item ``\textbf{Analyze} the documents for patterns and insights'' -- This prompt allows the LLM to apply any techniques to extract evidence.
    \item ``\textbf{Summarize} the main events described'' -- This prompt attempts to provide a concise description of the main events across the documents.
    \item ``\textbf{Extract} relevant entities (people, locations, etc.)'' -- This prompt uses the LLM to identify the most important entities across documents, e.g., to align with KG.
    \item ``\textbf{Classify} the relevant topics discussed'' -- This prompt tries to extract topics across documents to give an overview of what is in them.
    \item ``\textbf{Generate} potential questions that the documents raise'' -- This prompt helps analysts curate questions the LLM might be able to answer, using the LLM as its own evaluator.
    \item ``\textbf{List} analytic tasks to perform based on the documents'' -- This prompt gives inspiration for how to parse the documents given their structure.
    \item ``\textbf{Explain} concepts mentioned in the documents'' -- This prompt helps generate deeper, more focused answers to concepts listed by the user.
    \item ``\textbf{Answer} questions using the documents'' -- This prompt helps users structure questions for the LLM to answer.
    \item ``\textbf{Custom} prompt with open-ended input'' -- This prompt is blank; users write whatever they want.
\end{enumerate}

\begin{figure}[!t]
  \centering
  \includegraphics[width=\linewidth]{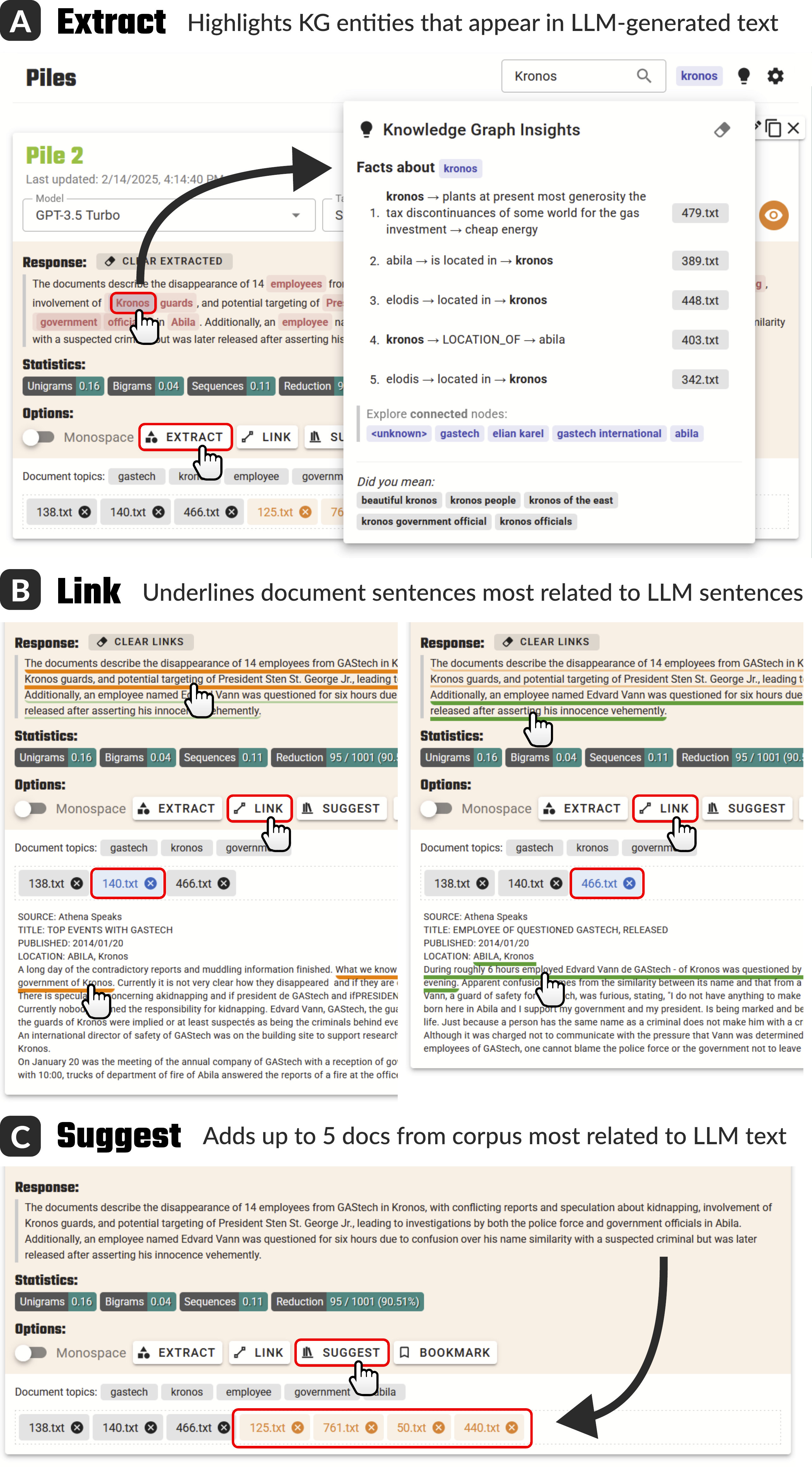}
  \caption{%
    Extract (\textbf{A}), Link (\textbf{B}), and Suggest (\textbf{C}) are buttons in piles that help analysts verify and contextualize evidence from the LLM and KG.
  }%
  \label{fig:validation_features}
\end{figure}

The KG fact list can also be directly traversed from a pile (\textbf{G4}).
The Extract button in piles (Fig.~\ref{fig:validation_features}A) automatically highlights entities from the KG that appear in LLM-generated text.
Clicking on a highlighted entity automatically searches for it in the KG fact list (Fig.~\ref{fig:piles_and_facts}B) and shows up to 5 related facts.
These facts can be traversed by directly clicking on entity names, allowing users to traverse the KG starting with facts related to their pile.

\medskip
\noindent\textbf{LLM \& KG validation features. }
Several features keep users in the loop of linking LLM- and KG-generated text with source documents, inspiring trust and increasing transparency throughout the sensemaking process.

The Extract button (Fig.~\ref{fig:validation_features}A) also helps with validation -- it can reveal limitations of the LLM like hallucinated terms (\textbf{G5}).
The Link button (Fig.~\ref{fig:validation_features}B) connects each sentence in the LLM response to the most related sentence in documents in a pile using a RAG-like approach.
The most similar sentence pairs are underlined and color-coded by the source document.
Links can be hovered/clicked to focus the opacity and more easily find pairs.
The Suggest button (Fig.~\ref{fig:validation_features}C) performs another similarity search of the entire LLM response in a pile against the entire document corpus, not just in the pile.
The top $5$ highest ranked documents are then automatically added to the pile, expanding evidence to corroborate the LLM response.
Only documents not already in the current pile out of the top $5$ are then added to the pile.
This avoids continuously adding documents to the pile that may be less relevant than what is already in the pile.
Finally, to help contextualize KG facts, the most connected KG entities, as well as semantically similar entities related to the currently searched term, can be clicked (Fig.~\ref{fig:piles_and_facts}B) to automatically populate the menu with more facts (\textbf{G6}).

\subsection{Usage Scenario}
To help envision how VisPile supports sensemaking, we \add{analyzed the KRONOS dataset ourselves using VisPile and formatted our findings as insights in a usage scenario}.
Consider a fictitious seasoned investigative analyst, Bob.
Bob has been hearing about a possible kidnapping, and they want to identify relationships between entities in the news for evidence of a collusion ring.
Every morning, $\approx1000$ daily news articles flood Bob's inbox.
With limited time to read every document, Bob likes using VisPile to enhance their sensemaking process going from documents to evidence.

\medskip
\noindent\textbf{Searching for and piling documents. }
Gaining situational awareness into a document corpus is tricky: What are the documents about? How are they related? Which documents are relevant?
To aid Bob in the foraging process, VisPile enhances searching, filtering, grouping, and sorting text documents.

Bob first filters the documents using known keywords like \textit{``kronos''} and groups documents by LLM-generated topic.
Then, Bob uses LLM-powered semantic search to rank the relevance of the remaining documents based on open-ended phrases like \textit{``reports related to kidnapping''} and \textit{``missing employees''}, including documents that may not explicitly mention these topics but are relevant (\textbf{G1}).
If needed, Bob can hover on documents in the documents list or click on the document blocks in a pile to expand the pile and read the source text at any time.
Bob then searches for a name, \textit{``edvard vann''}, in the KG and finds facts related to police investigations, with source documents that were missed by the LLM semantic search -- a potentially related group of documents (\textbf{G2}).
Once Bob finds relevant documents, they drag them into a pile and rename it to track their line of questioning.

\medskip
\noindent\textbf{Analyzing document piles. }
The piling process helped Bob quickly grasp key documents, entities, and events, organizing them into piles for deeper analysis.
However, with limited time, Bob cannot read all of the documents in piles to gather evidence.
Bob instead leverages the capabilities of LLMs and KGs directly in piles, helping them read less and reason more.

Bob requests a verbose summary using \textbf{summarize} to list the main events in their pile and get a lay of the land.
They follow up with \textbf{answer questions} to extract additional evidence from their pile (\textbf{G3}).
Customizing LLM tasks helps Bob get more details at the start that they can filter down over time (\textbf{G3}).
As Bob analyzes events related to the kidnapping, they hit the Extract button to link the LLM response with the KG.
A known business leader's name frequently appears, so Bob clicks on it to get more evidence from the KG (\textbf{G4}).
Bob discovers past police run-ins and continues clicking names to explore potential connections (\textbf{G4}).
They add connected documents to their pile and ask the LLM about the business leader's financial ties to the police.

\medskip
\noindent\textbf{Validating LLM \& KG results. }
With a growing set of evidence gathered from the LLM and KG, Bob needs to filter the noise and find signals to gain insight.

Bob first hits the Link button to map the LLM response sentences to source documents in their pile, uncovering new evidence of businesses that were never mentioned when prompting the LLM (\textbf{G5}).
Wondering if they missed any relevant documents on the financial ties of the businesses, Bob hits the Suggest button and sees several new documents appear that they haven't explored yet.
One mentions an interview, buried years ago and only today just coming to light, of a woman claiming the same businessman paid her hush money not to talk about the disappearance of her daughter (\textbf{G5}).

Bob continues exploring connected nodes in the KG to find evidence of relationships with the current suspect and the woman from the suggested documents (\textbf{G6}).
Several facts deep, the name of a business is listed as semantically related to the woman's news interview, yet it has not appeared in any KG facts thus far.
Curious, Bob searches in the KG for the business, finding several disconnected facts which corroborate the woman's story and implicate the local business leader (\textbf{G6}).
With these pieces of verified evidence, Bob alerts their superior that a potential connection has been found.
In tandem, the capabilities for validating and contextualizing LLM and KG evidence give Bob confidence that they have identified the right complex relationships hidden in their document corpus.

\section{Domain Expert Feedback}
\label{sec:expert_feedback}

Following \textcite{Sedlmair:2012:DesignStudyMethods}, we recruited $6$ professional analysts from the U.S. Intelligence Community (N$1-6$) to use VisPile and provide domain expert feedback.
All analysts have experience analyzing text documents using various visual analytics tools for gathering evidence and maintaining situational awareness.
Each analyst spent $60$ minutes freely exploring the KRONOS dataset (\cite{Whiting:2014:VAST2014Challenge}).
They were encouraged to give verbal feedback on usability and usefulness for supporting unclassified workflows.
We manually recorded their think-aloud feedback (\cite{Ericsson:1984:ProtocolAnalysis}) and inductively collected and organized high-level themes, discussing them amongst all authors (\cite{Boyatzis:1998:ThematicAnalysis}).
This process resulted in emergent feature usage, user workflows, and analytical trade-offs in a task-driven scenario.

\medskip
\noindent\textbf{Feature usage. }
The analysts generally liked features in VisPile that they felt could enhance their own sensemaking workflows.
Many analysts took advantage of the LLM and KG features to quickly explore alternative hypotheses.
N$1,2,5$ used the LLM similarity search for pairwise comparison of document sets.
N$1$ used Link to shortlist new documents to pile and explore, while N$3$ used topics to pile documents.
While \textit{``extract entities''}, \textit{``summarize events''}, and \textit{``analyze documents''} were slightly more commonly used, we observed many LLM tasks being tested.
For example, N$3$ used \textit{``generate questions''} to look for threads to seed their investigation: \textit{``I liked that the GPT can generate things for me. I didn't have to read 200 documents. I can get a summary or questions to ask about the documents without having to read them.''}
The KG was most useful for direct answers to questions, and both N$3$ and N$5$ used the KG for this reason.
N$3$ explains that \textit{``If someone wanted me to answer a specific question about the documents, I would go into a knowledge graph to find facts and find related documents.''}

\medskip
\noindent\textbf{Workflows. }
Analysts primarily tested features that helped them avoid reading, as N$5$ summarized: \textit{``There's a lot of features for not reading documents, and I want to use those features''}.
Every analyst started with larger piles and split them apart as they went.
N$1$ explained how LLMs and KGs can enhance domain-specific benefits of this sensemaking loop: \textit{``As an analyst, as I go through large amounts of data, I want to get a large amount of patterns and data that I can't keep in my head. I tend to start wider because if AI can help me, it helps give an overview much faster and if it hallucinates, I can drill down much faster and validate the information.''}
N$1$ also had a useful approach to staging LLM and KG tasks in this loop: \textit{``Extract entities sounds appealing early on because I can find entity names, summarize for events, classify topics... Later, I want to answer questions, list tasks, etc.''}
This may suggest a desire for tools to stage LLM and KG tasks and chain them together at different stages of analysis.

\medskip
\noindent\textbf{Challenges. }
Trust and understanding in the LLM and KG was a major concern for every analyst.
Many LLM limitations were encountered, such as hallucinating the name for an acronym for N$5$, giving inconsistent responses for N$6$, and missing a fact in the dataset for N$2$ (\textit{``I didn't know whether the information was in the documents, or if the problem was in the GPT''}).
Most had to warm up to trusting the LLM and KG, like N$4$: \textit{``The trust is never there for me at the start. I take everything with a grain of salt and then read more.''}
To overcome these issues, the Link button helped N$1,2,4$ perform fact-checking.
N$1$ had more confidence in using the LLM for tasks that it may be better at than humans: \textit{``I trust the LLM that it will be able to find information from a large dataset better than I will. I want to start with what the LLM says and go from there.''}
N$2$ used Suggest to help them find the answer to a question that they knew existed in the dataset, but that the LLM was not providing.
Traceability for both the KG and LLM are vital to realizing LLM- and KG-infused workflows.

\section{Conclusions and Future Work}
\label{sec:conclusion}

Using AI to automate and enhance sensemaking can enable new analysis capabilities that take us beyond what is currently possible in existing visual analytics systems.
Towards this goal, our work explores design opportunities to infuse LLMs and KGs into visual text analysis.
In collaboration with intelligence community experts, we developed VisPile, a visual analytics system for grouping documents, sensemaking, and validating evidence using LLMs and KGs.
Feedback from experts highlighted the usefulness of synergizing LLM and KG features, staging LLM and KG operations in a workflow, and using analytic provenance to enhance transparency.

Our work has several limitations.
\add{While the KRONOS dataset is widely used, it contains only short, plain-text documents less than $1000$ words, making the generalizability of our techniques to other, real-world datasets with longer documents unclear.}
We only tested OpenAI’s GPT-3.5 Turbo for prompting and KG generation; exploring other LLM sizes and context lengths could impact the quality of insights and the analyst experience.
We also did not perform entity matching or KG alignment across people, places, and events, which could enhance KG quality and analytic evidence.
\add{Similarly, factual accuracy might improve by exploring UI mechanisms for detecting and correcting errors during KG extraction. Although we explained each feature, the labeling of LLM settings like temperature and the pile validation buttons may have been unclear, confusing users. It is also unclear how limiting KG facts to the top five affected sensemaking, or whether showing all facts would improve understanding. While lists simplify interpretation, the absence of a node-link diagram may have limited analysts' ability to visualize the overall KG structure.}
Finally, prior research suggests generative AI may affect expert critical thinking (\cite{Lee:2025:GenAICriticalThinking}).
\add{For example, analyst feedback provided evidence to suggest that features in VisPile may help in building trust throughout the analysis process.}
However, more work is needed to study how VisPile affects sensemaking\add{ and trust-building} across tasks, datasets, and analyst groups.

\printbibliography

@book{United:2019:AIM,
    title={The AIM initiative: A strategy for augmenting intelligence using machines},
    author={ODNI},
    year={2019},
    publisher={ODNI}
}

@article{Sayler:2020:AINatSec,
    title={Artificial intelligence and national security},
    author={Sayler, Kelley M},
    journal={Congressional Research Service},
    volume={45178},
    year={2020}
}

@ARTICLE{Zhao:2025:LEVA,
  author={Zhao, Yuheng and Zhang, Yixing and Zhang, Yu and Zhao, Xinyi and Wang, Junjie and Shao, Zekai and Turkay, Cagatay and Chen, Siming},
  journal={IEEE Transactions on Visualization and Computer Graphics}, 
  title={LEVA: Using Large Language Models to Enhance Visual Analytics}, 
  year={2025},
  volume={31},
  number={3},
  pages={1830-1847},
  
}

@inproceedings{Pirolli:2005:SensemakingProcess,
    title={The sensemaking process and leverage points for analyst technology as identified through cognitive task analysis},
    author={Pirolli, Peter and Card, Stuart},
    booktitle={Proceedings of international conference on intelligence analysis},
    volume={5},
    pages={2--4},
    year={2005},
    organization={McLean, VA, USA}
}

@Article{Shipman:1999:IncrementalFormalism,
    author={Shipman, Frank M.
    and Marshall, Catherine C.},
    title={Formality Considered Harmful: Experiences, Emerging Themes, and Directions on the Use of Formal Representations in Interactive Systems},
    journal={Computer Supported Cooperative Work (CSCW)},
    year={1999},
    month={12},
    day={01},
    volume={8},
    number={4},
    pages={333-352},
    issn={1573-7551},
}

@article{Baker:2009:VisEnhancesSensemaking,
  title={Using visual representations of data to enhance sensemaking in data exploration tasks},
  author={Baker, Jeff and Jones, Donald and Burkman, Jim},
  journal={Journal of the Association for Information Systems},
  volume={10},
  number={7},
  pages={2},
  year={2009}
}

@INPROCEEDINGS{Kang:2011:VAIntelProcess,
    author={Kang, Youn-ah and Stasko, John},
    booktitle={2011 IEEE Conference on Visual Analytics Science and Technology (VAST)}, 
    title={Characterizing the intelligence analysis process: Informing visual analytics design through a longitudinal field study}, 
    year={2011},
    volume={},
    number={},
    pages={21-30},
    
}

@inproceedings{Mccolgin:2006:QAtoVA,
    title={From Question Answering to Visual Exploration},
    author={McColgin, David and Gregory, Michelle and Hetzler, Elizabeth and Turner, Alan},
    booktitle={Proceedings of the ACM SIGIR workshop on Evaluating Exploratory Search Systems, EESS 2006 Workshop},
    pages={47--50},
    year={2006},
    organization={Citeseer}
}

@INPROCEEDINGS{Kim:2019:TopicSifter,
    author={Kim, Hannah and Choi, Dongjin and Drake, Barry and Endert, Alex and Park, Haesun},
    booktitle={2019 IEEE Conference on Visual Analytics Science and Technology (VAST)}, 
    title={TopicSifter: Interactive Search Space Reduction through Targeted Topic Modeling}, 
    year={2019},
    volume={},
    number={},
    pages={35-45},
    
}

@INPROCEEDINGS{Stasko:2007:Jigsaw,
    author={Stasko, John and Gorg, Carsten and Liu, Zhicheng and Singhal, Kanupriya},
    booktitle={2007 IEEE Symposium on Visual Analytics Science and Technology}, 
    title={Jigsaw: Supporting Investigative Analysis through Interactive Visualization}, 
    year={2007},
    volume={},
    number={},
    pages={131-138},
    
}

@inproceedings{Endert:2012:SemanticInteraction,
    author = {Endert, Alex and Fiaux, Patrick and North, Chris},
    title = {Semantic interaction for visual text analytics},
    year = {2012},
    isbn = {9781450310154},
    publisher = {Association for Computing Machinery},
    address = {New York, NY, USA},
    booktitle = {Proceedings of the SIGCHI Conference on Human Factors in Computing Systems},
    pages = {473-482},
    numpages = {10},
    location = {Austin, Texas, USA},
    series = {CHI '12}
}

@inproceedings{Andrews:2010:SpaceToThink,
    author = {Andrews, Christopher and Endert, Alex and North, Chris},
    title = {Space to think: large high-resolution displays for sensemaking},
    year = {2010},
    isbn = {9781605589299},
    publisher = {Association for Computing Machinery},
    address = {New York, NY, USA},
    booktitle = {Proceedings of the SIGCHI Conference on Human Factors in Computing Systems},
    pages = {55-64},
    numpages = {10},
    location = {Atlanta, Georgia, USA},
    series = {CHI '10}
}

@Article{Endert:2014:HumanIsTheLoop,
    author={Endert, Alex
    and Hossain, M. Shahriar
    and Ramakrishnan, Naren
    and North, Chris
    and Fiaux, Patrick
    and Andrews, Christopher},
    title={The human is the loop: new directions for visual analytics},
    journal={Journal of Intelligent Information Systems},
    year={2014},
    month={12},
    day={01},
    volume={43},
    number={3},
    pages={411-435},
    issn={1573-7675},
}

@book{Berry:2010:TextMining,
    title={Text mining: applications and theory},
    author={Berry, Michael W and Kogan, Jacob},
    year={2010},
    publisher={John Wiley \& Sons}
}

@misc{i2:2024:AnalystNotebook,
	author = {{i2 Group}},
	date = {2024-08},
	publisher = {N. Harris Computer Corporation},
	title = {{Analyst's Notebook}},
	type = {software},
	version = {10.0.2},
}

@article{Srivastava:2023:BeyondImitationGame,
    title={Beyond the Imitation Game: Quantifying and extrapolating the capabilities of language models},
    author={Aarohi Srivastava and others},
    journal={Transactions on Machine Learning Research},
    issn={2835-8856},
    year={2023},
    note={Featured Certification}
}

@ARTICLE{Li:2024:KGChallenges,
    author={Li, Harry and others},
    journal={IEEE Transactions on Visualization and Computer Graphics}, 
    title={Knowledge Graphs in Practice: Characterizing their Users, Challenges, and Visualization Opportunities}, 
    year={2024},
    volume={30},
    number={1},
    pages={584-594},
    
}

@inproceedings{Cheng:2023:GPT4DataAnalyst,
    title = "Is {GPT}-4 a Good Data Analyst?",
    author = "Cheng, Liying  and
      Li, Xingxuan  and
      Bing, Lidong",
    editor = "Bouamor, Houda  and
      Pino, Juan  and
      Bali, Kalika",
    booktitle = "Findings of the Association for Computational Linguistics: EMNLP 2023",
    month = dec,
    year = "2023",
    address = "Singapore",
    publisher = "Association for Computational Linguistics",
    pages = "9496--9514",
}

@inproceedings{Zhang:2024:DataCopilot,
    title={Data-Copilot: Bridging Billions of Data and Humans with Autonomous Workflow},
    author={Wenqi Zhang and Yongliang Shen and Weiming Lu and Yueting Zhuang},
    booktitle={ICLR 2024 Workshop on Large Language Model (LLM) Agents},
    year={2024},
}

@inproceedings{Shaib:2023:LLMsForMultidocAnalysis,
    title = "Summarizing, Simplifying, and Synthesizing Medical Evidence using {GPT}-3 (with Varying Success)",
    author = "Shaib, Chantal  and
      Li, Millicent  and
      Joseph, Sebastian  and
      Marshall, Iain  and
      Li, Junyi Jessy  and
      Wallace, Byron",
    editor = "Rogers, Anna  and
      Boyd-Graber, Jordan  and
      Okazaki, Naoaki",
    booktitle = "Proceedings of ACL 2023 (Volume 2: Short Papers)",
    month = jul,
    year = "2023",
    address = "Toronto, Canada",
    publisher = "Association for Computational Linguistics",
    pages = "1387--1407",
}

@INPROCEEDINGS{Li:2024:LinkQ,
  author={Li, Harry and others},
  booktitle={2024 IEEE Visualization and Visual Analytics (VIS)}, 
  title={LinkQ: An LLM-Assisted Visual Interface for Knowledge Graph Question-Answering}, 
  year={2024},
  volume={},
  number={},
  pages={116-120},
  
}

@article{Liu:2023:Prompting,
    author = {Liu, Pengfei and Yuan, Weizhe and Fu, Jinlan and Jiang, Zhengbao and Hayashi, Hiroaki and Neubig, Graham},
    title = {Pre-train, Prompt, and Predict: A Systematic Survey of Prompting Methods in Natural Language Processing},
    year = {2023},
    issue_date = {September 2023},
    publisher = {Association for Computing Machinery},
    address = {New York, NY, USA},
    volume = {55},
    number = {9},
    issn = {0360-0300},
    journal = {ACM Comput. Surv.},
    month = {1},
    articleno = {195},
    numpages = {35},
}

@INPROCEEDINGS{Latif:2021:VisKonnect,
    author={Latif, Shahid and Agarwal, Shivam and Gottschalk, Simon and Chrosch, Carina and Feit, Felix and Jahn, Johannes and Braun, Tobias and Tchenko, Yanick Christian and Demidova, Elena and Beck, Fabian},
    booktitle={2021 IEEE Visualization Conference (VIS)}, 
    title={Visually Connecting Historical Figures Through Event Knowledge Graphs}, 
    year={2021},
    volume={},
    number={},
    pages={156-160},
    
}

@article{Hogan:2021:KGs,
    author = {Hogan, Aidan and others},
    title = {Knowledge Graphs},
    year = {2021},
    issue_date = {May 2022},
    publisher = {Association for Computing Machinery},
    address = {New York, NY, USA},
    volume = {54},
    number = {4},
    issn = {0360-0300},
    journal = {ACM Comput. Surv.},
    month = jul,
    articleno = {71},
    numpages = {37},
}

@ARTICLE{Yan:2025:KNowNEt,
    author={Yan, Youfu and Hou, Yu and Xiao, Yongkang and Zhang, Rui and Wang, Qianwen},
    journal={IEEE Transactions on Visualization and Computer Graphics}, 
    title={KNowNEt:Guided Health Information Seeking from LLMs via Knowledge Graph Integration}, 
    year={2025},
    volume={31},
    number={1},
    pages={547-557},
    doi={10.1109/TVCG.2024.3456364}
}

@article{Grootendorst:2022:BERTopic,
  title={BERTopic: Neural topic modeling with a class-based TF-IDF procedure},
  author={Grootendorst, Maarten},
  journal={arXiv preprint arXiv:2203.05794},
  year={2022}
}

@inproceedings{Lewis:2020:RAGLLM,
    author = {Lewis, Patrick and others},
    booktitle = {Advances in Neural Information Processing Systems},
    pages = {9459--9474},
    publisher = {Curran Associates, Inc.},
    title = {Retrieval-Augmented Generation for Knowledge-Intensive NLP Tasks},
    volume = {33},
    year = {2020}
}

@article{Sahoo:2024:PromptEngineering,
    publtype={informal},
    author={Pranab Sahoo and Ayush Kumar Singh and Sriparna Saha and Vinija Jain and Samrat Mondal and Aman Chadha},
    title={A Systematic Survey of Prompt Engineering in Large Language Models: Techniques and Applications},
    year={2024},
    cdate={1704067200000},
    journal={CoRR},
    volume={abs/2402.07927},
}

@INPROCEEDINGS{Whiting:2014:VAST2014Challenge,
    author={Whiting, Mark and Cook, Kristin and Grinstein, Georges and Liggett, Kristen and Cooper, Michael and Fallon, John and Morin, Marc},
    booktitle={2014 IEEE Conference on Visual Analytics Science and Technology (VAST)}, 
    title={VAST challenge 2014: The Kronos incident}, 
    year={2014},
    volume={},
    number={},
    pages={295-300},
    
}

@ARTICLE{Pan:2024:UnifyingLLMsKGs,
    author={Pan, Shirui and Luo, Linhao and Wang, Yufei and Chen, Chen and Wang, Jiapu and Wu, Xindong},
    journal={IEEE Transactions on Knowledge and Data Engineering}, 
    title={Unifying Large Language Models and Knowledge Graphs: A Roadmap}, 
    year={2024},
    volume={36},
    number={7},
    pages={3580-3599},
    
}

@book{Boyatzis:1998:ThematicAnalysis,
  title={Transforming Qualitative Information: Thematic Analysis and Code Development},
  author={Boyatzis, R.E.},
  isbn={9780761909613},
  lccn={97045405},
  series={Transforming Qualitative Information: Thematic Analysis and Code Development},
  year={1998},
  publisher={SAGE Publications}
}

@book{Ericsson:1984:ProtocolAnalysis,
  title={Protocol analysis: Verbal reports as data.},
  author={Ericsson, K Anders and Simon, Herbert A},
  year={1984},
  publisher={the MIT Press}
}

@inproceedings{Lee:2025:GenAICriticalThinking,
    author = {Lee, Hao-Ping and others},
    title = {The Impact of Generative AI on Critical Thinking: Self-Reported Reductions in Cognitive Effort and Confidence Effects From a  Survey of Knowledge Workers},
    booktitle = {Proceedings of the ACM CHI Conference on Human Factors in Computing Systems},
    year = {2025},
    month = {4},
    publisher = {ACM},
}

@article{Sedlmair:2012:DesignStudyMethods,
  title={Design study methodology: Reflections from the trenches and the stacks},
  author={Sedlmair, Michael and Meyer, Miriah and Munzner, Tamara},
  journal={IEEE transactions on visualization and computer graphics},
  volume={18},
  number={12},
  pages={2431--2440},
  year={2012},
  publisher={IEEE}
}

\end{document}